# Influence of apical oxygen on the extent of in-plane exchange interaction in cuprate superconductors

Y. Y. Peng[1], G. Dellea[1], M. Minola[2], M. Conni[1], A. Amorese[3], D. Di Castro[4], G. M. De Luca[5], K. Kummer[3], M. Salluzzo[5], X. Sun[6], X. J. Zhou[6], G. Balestrino[4], M. Le Tacon[2,7], B. Keimer[2], L. Braicovich[1,8], N. B. Brookes[3] and G. Ghiringhelli[1,8,*]

[1] *Dipartimento di Fisica, Politecnico di Milano, Piazza Leonardo da Vinci 32, I-20133 Milano, Italy.*
[2] *Max-Planck-Institut für Festkörperforschung, Heisenbergstraße 1, D-70569 Stuttgart, Germany.*
[3] *ESRF, The European Synchrotron, CS40220, F-38043 Grenoble Cedex, France.*
[4] *CNR-SPN and Dipartimento di Ingegneria Civile e Ingegneria Informatica, Università di Roma Tor Vergata, Via del Politecnico 1, I-00133 Roma, Italy*
[5] *CNR-SPIN, Complesso MonteSantangelo - Via Cinthia, I-80126 Napoli, Italy.*
[6] *Beijing National Laboratory for Condensed Matter Physics, Institute of Physics, Chinese Academy of Sciences, Beijing 100190, China*
[7] *Institute of Solid State Physics (IFP), Karlsruhe Institute of Technology, D-76021 Karlsruhe, Germany*
[8] *CNR-SPIN, Dipartimento di Fisica, Politecnico di Milano, Piazza Leonardo da Vinci 32, I-20133 Milano, Italy*

*Correspondence to: giacomo.ghiringhelli@polimi.it

**In high $T_c$ superconductors the magnetic and electronic properties are determined by the probability that valence electrons virtually jump from site to site in the $CuO_2$ planes, a mechanism opposed by on-site Coulomb repulsion and favored by hopping integrals. The spatial extent of the latter is related to transport properties, including superconductivity, and to the dispersion relation of spin excitations (magnons). Here, for three antiferromagnetic parent compounds (single-layer $Bi_2Sr_{0.99}La_{1.1}CuO_{6+\delta}$, double-layer $Nd_{1.2}Ba_{1.8}Cu_3O_6$ and infinite-layer $CaCuO_2$) differing by the number of apical atoms, we compare the magnetic spectra measured by resonant inelastic x-ray scattering over a significant portion of the reciprocal space and with unprecedented accuracy. We observe that the absence of apical oxygens increases the in-plane hopping range and, in $CaCuO_2$, it leads to a genuine 3D exchange-bond network. These results establish a corresponding relation between the exchange interactions and the crystal structure, and provide fresh insight into the materials dependence of the superconducting transition temperature.**

In copper-based high critical temperature superconductors the detailed electronic structure close to Fermi level and the short- and mid-range magnetic interactions are governed by the same physical parameters, namely the interatomic hopping integrals and the on-site Coulomb repulsion [1]. Therefore superconductivity, magnetism and charge density modulations are intimately related in cuprates, and any model for transport properties have to comply with the two other properties. Conversely, the determination of hopping parameters from magnetic measurements can help clarify the origin of superconductivity. The energy scale of spin excitations at the magnetic Brillouin zone boundary (250-400 meV) is mainly determined by the large nearest-neighbor Cu-Cu superexchange interaction ($J$), and makes spin-fluctuations good candidate for Cooper pairing in high $T_c$ superconductors[2]. This common property is easily traced back to the two-dimensional $CuO_2$ square lattice of all layered cuprates, where the Cu-O-Cu double bond has very similar length (3.8-3.9 Å) and angle (174°-180°) in all families, irrespective of the different out-of-plane structure. Attempts of correlating the value of $J$ with the critical temperature at optimal doping ($T_{c,max}$) did not reach consensus [3,4,5] because disorder, buckling and other properties can be more relevant than bare superexchange[6]. Indeed such large $J$ preserves short range in-plane antiferromagnetic correlation up to room temperature in undoped compounds, and well above optimal doping level in superconductors: the sharp and dispersing magnetic excitations measured by inelastic neutron scattering (INS)[7,8] or resonant inelastic x-ray scattering (RIXS)[9,10,11] in insulating compounds, survive in a damped, broadened form (paramagnon) throughout the superconducting phase diagram[9,12,13,14].

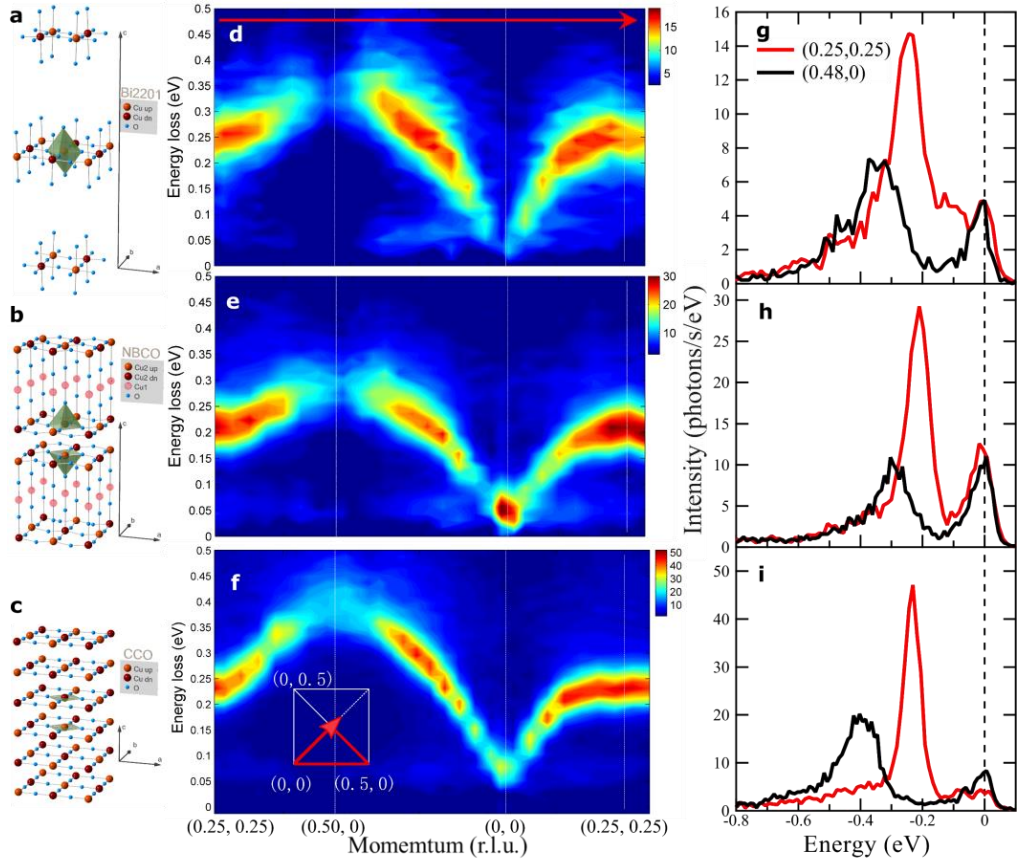

**Figure 1: In-plane momentum dependence of the magnetic excitations of antiferromagnetic layered cuprates measured by RIXS at the Cu $L_3$ resonance. a** to **c,** Partial crystalline structures of $Bi_2Sr_{2-x}La_xCuO_6$, $NdBa_2Cu_3O_{6+x}$ and $CaCuO_2$. **d** to **f,** Spin wave dispersion of heavily underdoped $Bi_2Sr_{2-x}La_xCuO_6$ ($p$=0.03), undoped $NdBa_2Cu_3O_{6+x}$ and $CaCuO_2$ respectively along the high symmetry momentum trajectory indicated in the inset of **f**. Elastic peaks were subtracted for a better visualization of the low energy features. **g** to **i,** Raw spectra at high symmetry points $q_{\parallel}$=(0.25, 0.25) (red) and $q_{\parallel}$=(0.48,0) (blue) belonging to the antiferromagnetic Brillouin zone boundary: the comparison highlights the differences in intensity, energy and width between the two points, and the actual elastic contribution.

However longer range electronic (and spin) correlations are those that determine the exact shape of the spin excitation dispersion, which is peculiar to each family of cuprates. It was shown by Coldea et al.[7] that in $La_2CuO_4$ (LCO) the magnon dispersion can be adequately reproduced by considering a cyclic exchange beyond the nearest-neighbor Heisenberg term. It is common to tie down those longer range effective exchange integrals by expressing them in terms of Hubbard model parameters, i.e. the nearest neighbor (Cu to Cu) hopping integral $t$ and the Coulomb repulsion $U$, where next-nearest neighbor hopping, described by multiple jumps, is proportional to higher powers of $t$. This approximation can reproduce a sizable energy dispersion along the magnetic Brillouin zone boundary (MBZB), but it usually leads to unphysically small $U$, and cannot fully account for the departure from a sine law of $\omega(q_{\parallel})$ along the (1,0) direction[7,11,15]. The bare inclusion, in the one band Hubbard model, of next-nearest-neighbor hopping parameters $t'$ and $t''$ further improves the fitting to experimental results for $La_2CuO_4$, $Sr_2CuO_2Cl_2$ and Y-doped Bi2212 (ref. [16]) but cannot be easily related to the other properties of those materials. Pavarini et al.[17], by considering a Hubbard model including $Cu4s$, $Cu3d_{z^2}$ and apical-$O2p_z$ (in addition to the usual in-plane $Cu3d_{x^2-y^2}$ and $O2p_{x,y}$), suggested larger intralayer hopping range leads to higher $T_{c,max}$.

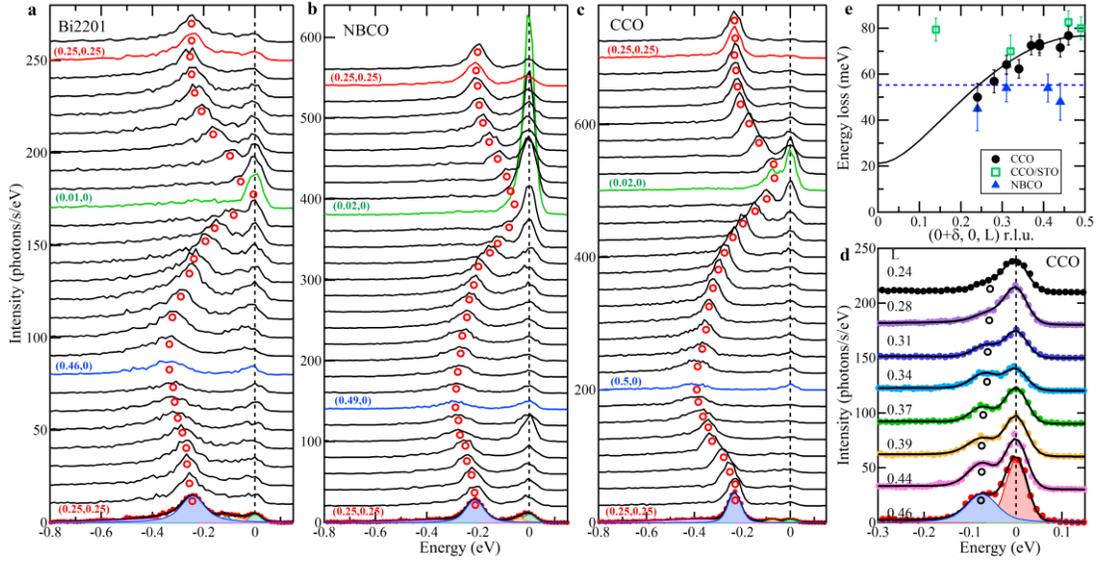

**Fig. 2: Spectral fitting and three-dimensional dispersions of magnetic excitations in layered cuprates.** The raw RIXS spectra for Bi2201 (**a**), NBCO (**b**) and CCO (**c**) measured at equally spaced positions along in-plane high symmetry directions described in the inset of Fig. 1**f**. Each spectrum is shifted vertically for clarity. Examples of the five (three) peaks decomposition described in Supplementary text and Figs. S3 (Figs. S5) are shown in the bottom spectra. Circles denote the peak positions of spin excitations determined by fitting. **d,** In CCO, dispersion of the peak at zone center along the normal direction of the $CuO_2$ planes: (0.015,0,$L$) trajectory for $L$=0.24-0.46. **e,** Summary of the ($\delta$,0,$L$) dispersions of magnetic excitations for CCO ($\delta$=0.015), superlattice CCO/STO ($\delta$=0.05) and NBCO ($\delta$=0.02). The $L$ values are reduced to symmetry-equivalent reciprocal-space points within the 0-0.5 interval. The black line (blue dashed line) is the dispersion calculated with a $J_\perp$=4.3 meV (6.2 meV) for CCO (NBCO). The error bars reflect the accuracy of the fitting procedure as detailed in Supplementary Information.

Here, by exploiting the new experimental capabilities of Cu $L_3$ edge RIXS, we determine the magnon dispersion of three antiferromagnetic cuprates over the whole magnetic antiferromagnetic Brillouin zone, we analyze the results within the Heisenberg model with extended exchange terms and relate them to the crystalline structure. Figure 1 provides a comprehensive overview of our experimental results for three materials differing by the number of apical oxygens per Cu atom: single layer $Bi_2Sr_{0.99}La_{1.1}CuO_{6+\delta}$ (Bi2201, $p$=0.03) (ref. [18]), bi-layer $Nd_{1.2}Ba_{1.8}Cu_3O_6$ (NBCO) (ref. [19]) and infinite-layer $CaCuO_2$ (CCO) (ref. [20]). RIXS spectra were measured at equally spaced points in the in-plane reciprocal space, along the (¼,¼)→(½,0)→(0,0)→(0.30, 0,30) path (Fig. 1f, inset), fully representative of the first magnetic Brillouin zone. The out-of-plane momentum $L\neq0$ changes along the path and differs from sample to sample. Bi2201 is constituted by two $CuO_2$ planes per unit cell with (½ ,½) stacking offset, each with two symmetrical apical O atoms per Cu site highlighted by the elongated octahedron in Fig. 1a. The Cu-O sheets are very distant from each-other, resulting in a very weak inter-planar magnetic coupling. On the contrary, in NBCO the two $CuO_2$ planes per unit cell are only 3.2 Å apart, and only one apical O per in-plane Cu contributes to the pyramid depicted in Fig. 1b. In CCO the absence of apical ligands allows the infinite, compact stacking of $CuO_2$ layers, with similar separation as in NBCO (Fig. 1c). The color maps of Figs. 1d-1f show important differences, beyond the overall similar dispersion shape, intensity and energy scale. Near the point (0,0) in the 2D reciprocal space, spin-wave excitations have vanishing energy and intensity in Bi2201, while they have nonzero energy and intensity in NBCO and CCO. In NBCO this corresponds to the gapped branch at (½,½) seen in YBCO with INS (refs [21,22]) and is due to the antiferromagnetic coupling between the adjacent $CuO_2$ planes. The other branch, expected to go to zero at (0,0), is not visible because its intensity vanishes while approaching zone center. The nonzero spin wave energy of CCO near (0,0) is due to the interlayer magnetic coupling too. In addition, the infinite-layer structure gives rise to a long range out-of-plane spin-correlation as demonstrated below. For all samples the dispersion along the (¼,¼)→(½,0) line is larger than previously found in $La_2CuO_4$ (LCO), $Sr_2CuO_2Cl_2$ (SCOC) and Bi2212; and in CCO it is almost double than that in NBCO and Bi2201. In (½,0) the width of the magnon peak is clearly larger than in (¼,¼), well beyond the instrumental resolution (see Fig. 1g to 1i). This common feature was observed earlier in several square-lattice AF systems[8,15,23] and discussed in terms of possible coupling to higher energy $\Delta S=½$ "spinon" excitations[24].

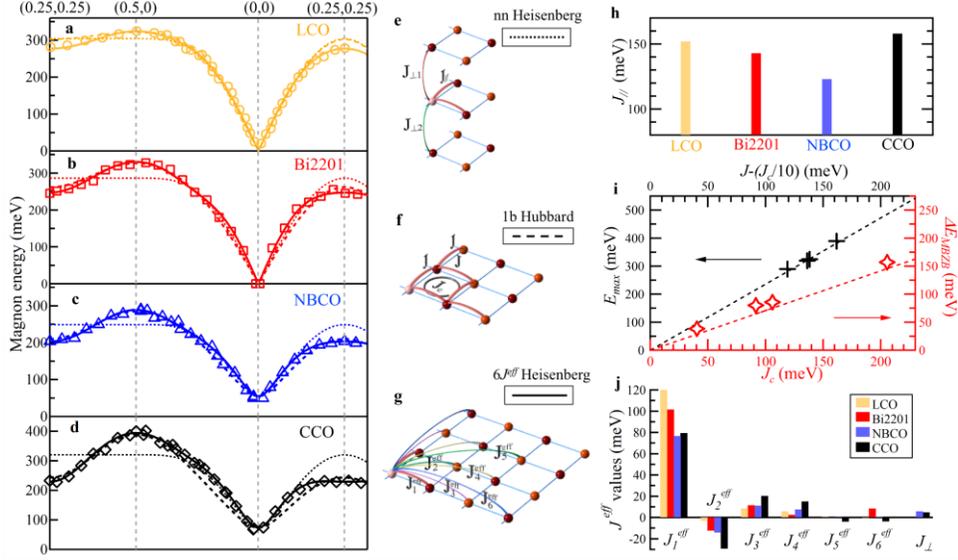

**Fig. 3: Dispersion of the spin excitations and comparison to model calculations. a** to **d,** Experimental magnon dispersion along the high symmetry direction in antiferromagnetic Bi2201, NBCO, CCO (RIXS) and LCO (INS) (ref.15), fitted using the nearest-neighbor Heisenberg (nnH) model (dotted line), the one band Hubbard (1bH) model (dashed line) and the phenomenological linear spin-wave Heisenberg model with six nearest-neighbor coupling parameters ($6J^{eff}$) (thick line). The error bars are smaller than symbol dimensions. **d** to **g,** Schematics of the nnH, 1bH and $6J^{eff}$ Heisenberg models respectively. **h,** The $J_{\parallel}$ values determined from the nnH model. **i,** Linear dependence of $E_{max}=E(½,0)$ (left/top axis, black symbols) and of $\Delta E_{MBZB}$ (right/bottom axis, red) vs. the one-band Hubbard model parameters $J$ and $J_c$ as defined in **f**; the lines follow equations (1,2) of the text. **j,** Effective parameters of the phenomenological spin wave model based on six in-plane $J^{eff}$ parameters as defined in **g**, and $J_{\perp}$ for NBCO and CCO.

The raw spectra shown in Fig. 2 were decomposed into a resolution-limited elastic peak, a resolution-limited phonon excitation, a weak resolution-limited bi-phonon excitation [25], an intense single-magnon peak and the bi-magnon tail (details in Supplementary Information). Besides the large dependence on the in-plane wave vector for all samples, the magnetic peak of CCO disperses also normally to the planes, along (0.015,0,$L$) (Fig. 2d). To understand this long range out-of-plane order we have grown and measured the $(CCO)_{3uc}/(STO)_{2uc}$ superlattice where the AF structure is made inherently 2D by the intercalation of $SrTiO_3$ sheets to form a sort of artificial tri-layer cuprate. Consequently the zone-center gap is $L$-independent in $(CCO)_{3uc}/(STO)_{2uc}$ superlattice, as in NBCO but at variance from pure CCO (Fig. 2e). Interestingly, although the in-plane dispersion is the same in CCO and CCO/STO superlattice (Fig. S7), superconductivity has been realized in the CCO/STO superlattice[26], but not in the perfect infinite-layer[27].

The magnon peak energy dispersions of our samples are presented in Fig. 3a-d together with that of LCO measured by INS (ref.15). The experimental points are fitted with 3 models of increasing complexity and exchange integral spatial extent. Following ref. [28] we initially fitted all magnon dispersions with a simple nearest-neighbor Heisenberg model based on one effective in-plane exchange $J_{\parallel}$ parameter and two interlayer couplings $J_{\perp 1}$ and $J_{\perp 2}$, neglecting the magnetic anisotropy (Fig. 3e). The values of $J_{\parallel}$ are shown in Fig. 3h and do not follow an obvious trend versus the number of apical oxygens. The model correctly accounts for the $L$-dependence (independence) of the spin gap and gives $E(0,0,1/2) \simeq 2\sqrt{2J_{\parallel}J_{\perp}}$ ($2\sqrt{J_{\parallel}J_{\perp}}$) in CCO (NBCO). Namely we find $J_{\parallel}(J_{\perp}) = 158$ meV (4.3 meV) for CCO and 123 meV (6.2 meV) for NBCO, which is comparable to 125±5 meV (11±2 meV) found for YBCO with INS (refs 21,22).

Although the simple nearest-neighbor Heisenberg model provides reasonable estimations of $J_{\parallel}$ and $J_{\perp}$, the large experimental dispersion along the MBZB calls for a better description of the in-plane magnetic interactions. Therefore we fitted our experimental results with the one-band Hubbard model with nearest-neighbor hopping $t$ as in Ref. 7, but keeping the same definition of $J_{\perp}$ for NBCO and CCO.

We replaced $J_\parallel$ by three in-plane magnetic interactions up to second nearest-neighbors that are ultimately expressed by only two independent parameters, $J$ and $J_c$ (or $t$ and $U$) (Fig. 3f). There $J = 4t^2/U - 24t^4/U^3$ is the nearest neighbor Cu-O-Cu superexchange integral, and $J_c = 80t^4/U^3$ is the so called "ring exchange" involving the 4 atoms of a plaquette. It can be shown that the maximum of magnon energy

$$E_{max} = E\left(\tfrac{1}{2}, 0, L\right) = 2Z_c\left(J - \tfrac{J_c}{10}\right) \quad (1)$$

is mainly set by $J$, whereas $J_c$ determines the energy dispersions along the zone boundary,

$$\Delta E_{MBZB} = E\left(\tfrac{1}{2}, 0, L\right) - E\left(\tfrac{1}{4}, \tfrac{1}{4}, L\right) = \tfrac{3}{5} Z_c J_c \quad (2)$$

where $Z_c$=1.18 is a renormalization factor accounting for quantum fluctuations[29]. The fitting values are listed in Table I and summarized in Fig. 3i, together with the trend lines following equations (1) and (2). The fitting curves (dashed lines in Fig. 3a-d) are very satisfactory overall. However the corresponding Hubbard model parameters imply $U<6t$ for all samples, except LCO: as a consequence $U$ is much smaller than that obtained from ARPES (ref. [30]), and the large-$U$ hypothesis at the basis of the Hubbard model is not fulfilled.

This inadequacy of the one-band Hubbard model with only nearest-neighbor hopping in accounting for the details of magnon dispersion in cuprates is a longstanding issue that has led to the extension of the model with the second nearest neighbor direct hopping parameters $t'$ and $t''$ (refs [31],[32]). Preferring a more phenomenological approach, we have simply relaxed the fitting constraints, abandoning the Hubbard model and adding longer range effective exchange terms up to 3$^{rd}$ nearest neighbors[33] as illustrated in Fig. 3g; we tested also other parameter sets restricted to a shorter range with no significant improvement with respect to the Hubbard model results. The resulting dispersion is given in Fig. 3a-d by the solid lines, which are considerably closer to the experimental points than the dashed ones: the fitting improves particularly at the intermediate $q_\parallel$ values, and the residual error is 2 to 3 times smaller. The 6$J^{eff}$ parameters are listed in Supplementary Table S3 and plotted in Fig. 3j. The main result is that the negative (ferromagnetic) exchange across the diagonal of the plaquette ($J_2^{eff}$), and the positive (antiferromagnetic) coupling with next nearest-neighbor along the edge ($J_3^{eff}$) both grow from LCO to Bi2201, to NBCO, to CCO, indicating an increasing importance of longer range interactions.

In summary, the fittings with the two models going beyond nearest neighbor interaction have shown that "less" apical oxygens imply larger zone boundary dispersion $\Delta E_{MBZB}$, and larger dispersion require bigger long-range exchange parameters ($J_c$, or $J_2^{eff}$, $J_3^{eff}$); conversely we confirm that $E_{max}$, the maximum energy at (½,0), and the related nearest-neighbor exchange parameters ($J_\parallel$, $J$, $J_1^{eff}$) are not univocally related to apical oxygens. To complete this phenomenological analysis we look also at the ligand field excitations, easily accessible by RIXS, and known to be strongly influenced by the local coordination of the Cu ion[34]. In particular, in the spectra of Fig. 4a we highlight the energy of the $3d_{z^2}$ orbital excitation ($E_z^2$) as obtained from the $dd$-excitation region of the RIXS spectra. We find that $E_z^2$ is positively correlated with $\Delta E_{MBZB}$, as illustrated in the bottom panel of Fig. 4b. Here closer apical oxygens are known to reduce the split-off energy of the $3d_{z^2}$ orbital and to increase its weight in the ground state[34]: it appears clearly that apical oxygens tend to localize charge in the $3d_{z^2}$ orbitals thus producing an effective screening by the polarizable charge reservoir layer[35].

Then the crucial question arises: how much are apical oxygens influencing superconductivity in cuprates? Theory and experiments are eventually pointing at a consistent answer. The relation between $\Delta E_{MBZB}$ and the maximum of $T_c$ are plotted in the top panel of Fig. 4b. This relation can be explained by the decrease of long range hopping integrals due to apical oxygens as originally stated by Pavarini et al.[17], who pointed out that the maximal $T_c$ in each family of cuprates scales with the next-nearest-neighbor hopping, which grows with the distance of apical oxygen to the $CuO_2$ plane. Similarly, we can define a range parameter $r = \left(\sqrt{|J_2^{eff}|} + \sqrt{J_3^{eff}}\right)/\sqrt{J_1^{eff}}$, assuming $J^{eff} \propto t^2$, and plot $r$ against $T_{c,max}$ in the inset of Fig. 4b. It shows $T_{c,max}$ tends to increase with $r$, in good qualitative agreement with Fig. 5 in Ref. 17. Our phenomenological findings are also in striking agreement with the theoretical works by Sakakibara et al.[36,37,38], who showed that the $3d_z^2$ orbital contribution to the Fermi surface is detrimental to $d$-wave superconductivity. The previous experimental support to their conclusions had come from ARPES, as the Fermi surface is shaped by the relative weight of next-nearest neighbor hopping integrals ($t'$,$t''$) to the first neighbor one ($t$). Our work establishes now a direct relation to the theory, as the link between exchange parameters and hopping integrals is straightforward and RIXS can be used on all cuprate families. Our results suggest looking directly into the role of longer-range exchange interactions, which may change the kinematics of the spin fluctuation mediated pairing interaction. However this goes beyond the present work and has to be addressed in future calculations in the spirit of Ref. 39.

**Table 1.** The one band Hubbard model parameters resulting from the least-square fitting of the Bi2201, NBCO and CCO (present work) and LCO (ref. 15) spin wave dispersion experimental data and leading to the dashed curves in Fig. 3a-d. $n$ and $z_{Cu-O}$ are the number of apical oxygens and their distance in Å, respectively. The other parameters $E_{max}$, $\Delta E_{MBZB}$, $J$, $t$, $U$ are expressed in meV.

|  | $n$ | $z_{Cu-O}$ | $E_{max}$ | $\Delta E_{MBZB}$ | $J$ | $J_c$ | $J_\perp$ | $t$ | $U$ | $U/t$ |
|---|---|---|---|---|---|---|---|---|---|---|
| LCO | 2 | 2.42 | 324.8 | 38.4 | 140.2 | 40.5 | - | 330.4 | 2865 | 8.70 |
| Bi2201 | 2 | 2.58 | 326.3 | 80.2 | 147.5 | 91.6 | - | 270.3 | 1670 | 6.17 |
| NBCO | 1 | 2.38 | 289.2 | 85.8 | 129.7 | 106.0 | 7.1 | 222.8 | 1229 | 5.52 |
| CCO | 0 | - | 400.1 | 157.2 | 182.3 | 205.6 | 6.4 | 297.1 | 1447 | 4.88 |

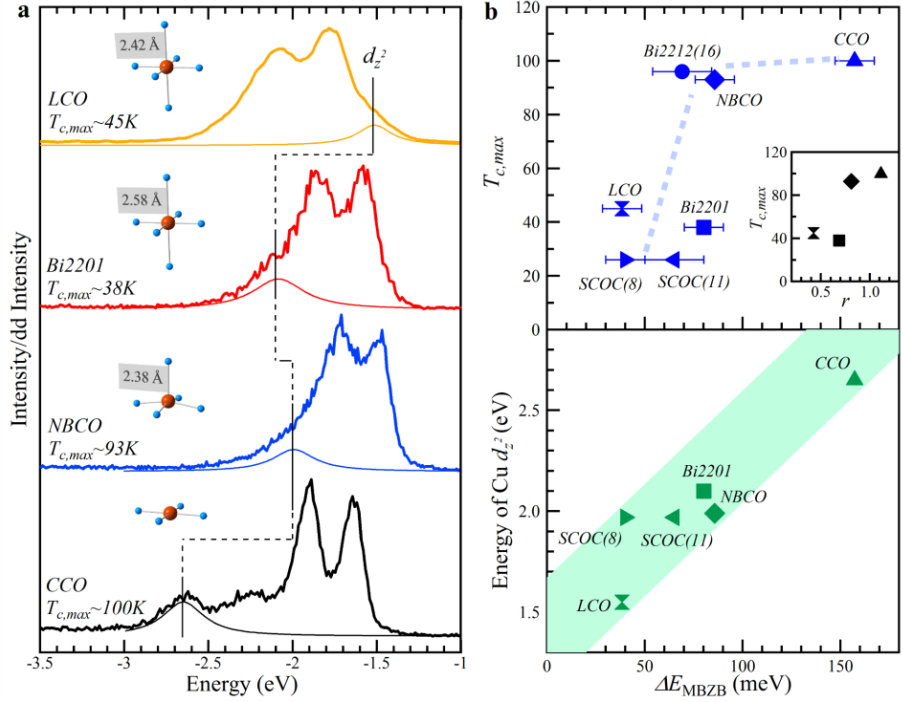

**Fig. 4: Phenomenological relation of the spin-wave dispersions with ligand field $d_z^2$ orbital excitations and $T_{c,max}$. a,** Normalized RIXS spectra, measured at $q_{//} \sim (0.48, 0)$ with π-polarization, in the $dd$ excitation region: the $d_z^2$ orbital components are shown as thin lines. The local Cu-O coordination is sketched in the inset. **b,** Correlation of the maximum critical temperature ($T_{c,max}$) (top) and the energy of Cu $d_z^2$ (bottom) with $\Delta E_{MBZB}$. Data of $Sr_2CuO_2Cl_2$ (SCOC) (refs 8,11) and $Bi_2Sr_2YCu_2O_8$ (Bi2212) (ref. 16) are also included. The inset shows $T_{c,max}$ plotted against the range parameter $r$, defined in the text. The $T_{c,max}$ values follow Ref. 6 except CCO (ref. 27). The lines highlight trends and are not intended to propose mathematical relationships.

Multi-orbital Hubbard model calculations and our present work suggest that an ideal hole-doped cuprate superconductor would maximize the long range hopping and minimize the $3d_z^2$ character of the Fermi surface. These properties are favored by structures with less and farther apical oxygens (larger crystal field splitting), which are determined also by Madelung potential difference between in-plane and apical oxygens[37], a difference traceable back to the details of the charge reservoir structure and chemical composition. Noticeable exceptions remain, like Bi2201, whose $T_{c,max}$ = 38 K is too low considering that the spatial extent of its hopping (exchange) integrals is comparable to that of single layer Tl2201 and Hg1201, and bi-layer NBCO/YBCO and Bi2212, all with $T_{c,max} > 90$ K; this exception has been explained by stronger out-of-plane disorders in Bi2201 (refs 6,40) and its $T_{c,max}$ may be considerably improved if the disorder is reduced. Therefore we can retain the message that *less apical is better for superconductivity*. Unfortunately infinite layer compounds, like CCO and (SrCa)$CuO_2$, include apical oxygens whenever hole doped (and superconducting). And in CCO/STO superlattices it has been demonstrated that only the interface layer, with apical oxygens, can be effectively doped to optimal level or above, while the inner layers remain underdoped[20,26]. Tri-layer compounds like Bi2223 and Tl2223 indeed reach very high $T_c$, because the inner $CuO_2$ layer has no apical, but its effective doping is also lower than nominal[41]. This explains why tri-layer cuprates usually have the largest $T_{c,max}$, as they provide an ideal trade-off between long range in-plane hopping and efficient inter-layer charge transfer.

## Methods
### Sample preparation.
Single crystals of $Bi_2Sr_{0.99}La_{1.1}CuO_{6+\delta}$ (Bi2201, $p$=0.03) were grown by the travelling solvent floating zone method. The sample growth and characterization methods have been reported previously[18]. The samples were cleaved out-of-vacuum to expose a fresh surface. $Nd_{1.2}Ba_{1.8}Cu_3O_6$ (NBCO) films were deposited by high oxygen pressure diode sputtering on $SrTiO_3$ (100) single crystals. After deposition the undoped NBCO film was obtained by annealing in Argon atmosphere (10 mbar) for 48 hours the as-grown $Nd_{1.2}Ba_{1.8}Cu_3O_7$ films[19]. The $CaCuO_2$ films and $CaCuO_2/SrTiO_3$ superlattice were grown by pulsed laser deposition (KrF excimer laser, $\lambda$ = 248 nm) on 5x5 mm$^2$ $NdGaO_3$ (110) substrate[20, 26]. Two targets, with $CaCuO_2$ and $SrTiO_3$ nominal composition, mounted on a multitarget system, were used. The STO target is a commercial crystal obtained from Crystal, GmbH. The CCO target was prepared by standard solid state reaction, according to the following procedure: stoichiometric mixtures of high-purity $CaCO_3$ and $CuO$ powders were calcined at 860 °C in air for 24 h, pressed to form a disk, and finally heated at 900 °C for 12 h. The substrate was placed at about 3 cm from the targets on a heated holder and its temperature during the deposition of the films was $T \approx$ 600 °C. The deposition chamber was first evacuated down to P~$10^{-5}$ mbar and then oxygen atmosphere at a pressure of about $5\times10^{-2}$ mbar was used for the growth. At this pressure, both CCO films and CCO/STO superlattice are insulating. At the end of the deposition, the films were cooled down at the growth pressure. The CCO films thickness is about 100 unit cells (~320 Å). The superlattice CCO/STO is formed by 20 repetitions of the "supercell" $CCO_3/STO_2$, made by 3 unit cells of CCO and 2 unit cells of STO. Reciprocal lattice units (r.l.u.) were defined using the lattice constants $a = b$ = 3.86 Å and $c$ =24.4 Å for Bi2201, $a = b$ = 3.84 Å and $c$ =11.7 Å for NBCO, $a = b$ = 3.85 Å and $c$ =3.2 Å (17.6 Å) for CCO (superlattice CCO/STO).

### RIXS measurements.
The RIXS measurements were performed at the new beam line ID32 of ESRF (The European Synchrotron, France) using the new high-resolution ERIXS spectrometer. ERIXS allows, for the first time in the soft x-ray range, experiments with complete sample orientation control and the possibility of changing in a continuous way the scattering angle of 2θ (50° to 150°) as shown in Supplementary Fig. S1a. The resonant conditions were achieved by tuning the energy of the incident x rays to the maximum of the Cu $L_3$ absorption peak, around 931 eV. The total instrumental band width (BW) of 55 meV at 931 eV (resolving power = 17,000) has been obtained with 15 micron exit and entrance slit on the monochromator, 4 micron spot size on the sample, the 800 lines mm$^{-1}$ grating of the monochromator, the 1400 lines mm$^{-1}$ grating of the spectrometer, and the single photon detection mode in the Princeton 2048×2048 13.5 micron pixel detector, cooled at -110°C by liquid nitrogen. The samples were cooled at 35 K, and were mounted on the 6 axis in-vacuum Huber diffractometer/ manipulator. The instrumental BW was measured as FWHM of the non-resonant diffuse scattering from polycrystalline graphite (carbon-tape). The typical size of the Brillouin zone in cuprates is 0.81 Å$^{-1}$ (0.5 r.l.u.) and the maximum total momentum transfer is 0.91 Å$^{-1}$ (2θ =150°), which allows one to cover the whole first Brillouin zone along the [100] or [010] direction (Supplementary Fig. S1b). In the specular geometry by changing the scattering angle 2θ one can measure the [001] direction at Brillouin zone center (Supplementary Fig. S1b). In-plane dispersion was measured with fixed 2θ = 149°. The exact position of the elastic (zero energy loss) line was determined by measuring, for each transferred momentum, a non-resonant spectrum of silver paint or carbon-tape. All data were obtained with π incident polarization (parallel to the scattering plane) to maximize the single magnon signal[10,42]. Each spectrum is the result of 15 or 20 minutes total accumulation (sum of individual spectra of 30 seconds).

## Acknowledgments
This work was supported by MIUR Italian Ministry for Research through project PIK Polarix. M. M. was supported by the Alexander von Humboldt Foundation. The authors acknowledge insightful discussions with Ole Andersen, Emanuele Dalla Torre, Tom Devereaux, Carlo Di Castro, Marco Grilli and Krzysztof Wohlfeld. The experimental data were collected at the beam line ID32 of the European Synchrotron (ESRF) in Grenoble (F) using the ERIXS spectrometer designed jointly by the ESRF and Politecnico di Milano.